\documentclass[aps,preprint]{revtex4}
\usepackage{amsmath}
\usepackage{bm}

\newcommand{\va}{\mathbf{a}}
\newcommand{\vb}{\mathbf{b}}

\newcommand{\ve}{\mathbf{e}}

\newcommand{\vvr}{\mathbf{r}}

\newcommand{\ha}{\mathbf{\hat a}}
\newcommand{\hb}{\mathbf{\hat b}}

\newcommand{\hr}{\mathbf{\hat r}}

\newcommand{\halb}{\frac{1}{2}}

\begin{document}

\title{Multipole expansions in four-dimensional hyperspherical harmonics} 

\author{A.~V.~Meremianin}
\email{avm@mpipks-dresden.mpg.de}

\affiliation{Max-Planck-Institute for the Physics of Complex Systems,
  N\"othnitzer Str. 38, 01187, Dresden, Germany}
\affiliation{REC-010, Voronezh State University, 394006, Voronezh, Russia}
\date{\today}

\begin{abstract}
The technique of vector differentiation is applied to the problem of the
derivation of multipole expansions in four-dimensional space.
Explicit expressions for the multipole expansion of the function 
$r^n C_j (\hr)$ with $\vvr=\vvr_1+\vvr_2$ are given in terms of tensor
products of two hyperspherical harmonics depending on the unit vectors $\hr_1$
and $\hr_2$. 
The multipole decomposition of the function $(\vvr_1 \cdot \vvr_2)^n$ is
also derived.
The proposed method can be easily generalised to the case of the space with
dimensionality larger than four.
Several explicit expressions for the four-dimensional Clebsch-Gordan
coefficients with particular values of parameters are presented in the closed
form.
\end{abstract}


\maketitle


\section{Introduction}
\label{sec:introduction}

The quantum angular momentum theory has proved to be a very efficient tool in
many physical problems.
Among the basis objects of that theory are the spherical harmonics which are
the solution of the angular part of the Laplace equation.
For spaces with dimensionality larger than three, the eigenfunctions of the
angular part of the Laplacian are often called ``hyperspherical harmonics''
(HSH).
From the point of view of quantum mechanics HSH are eigenfunctions of the
total angular momentum operator.
That explains why HSH have found wide use in the theory of quantum
few-particle systems.

The theory of three-dimensional spherical harmonics is well developed 
since J.C.~Maxwell's work on electromagnetic theory.
In particular, the use of spherical harmonics allows one to simplify the
calculation of various angular integrals.
However, before such an integral can be calculated, one has to re-write the
integrand functions in terms of spherical harmonics.
The decomposition of a some function in terms of spherical harmonics is
called the multipole expansion.
There is a large number of known multipole expansion formulas for different
functions in three-dimensional space, see
e.g.~\cite{Varsh,sack64:_three_dimen_addit_theor_arbit}.

The expansion formulas in many-dimensional space are less known.
The multipole expansion for the exponential function
$\exp(i\, \va \cdot \vb)$ has been derived in \cite{avery85jmp:_hyp_spher}.
This exponential function is a scalar and its multipole decomposition is
equivalent to the expansion in terms of Gegenbauer polynomials.
In the paper \cite{avery85jmp:_hyp_spher} also the differential method for
deriving the multipole expansions of scalar functions of the kind
$f(|\vvr_1-\vvr_2|)$ has been developed.

Generally, the explicit expressions for the coefficients of multipole
expansions (so-called ``multipole coefficients'') can be obtained using the
orthogonality properties of spherical harmonics.
This means that the multipole coefficients are determined by calculating the
overlap integrals between the function to be expanded and the corresponding
HSH.
It is clear that in spaces with higher dimensions those overlap integrals will
become more and more complicated.

The main goal of the present paper is to develop the differential technique
for the derivation of multipole expansions.
In the three-dimensional space such a technique is already known
\cite{manakov02:_multip}.
In the present paper, the multipole expansions in four-dimensional space are
considered.
The study of multipole expansions in four-dimensional space is motivated
mainly by the fact that four-dimensional spherical harmonics represent the wave
function of the hydrogen atom in momentum space \cite{fock35:_o4}.
Such harmonics are also used as Sturmian basis functions in many-centre
Coulomb problems, see 
e.g.~\cite{avery04:_shibuya_wulfman_fock,aquilanti01:_rev_hypsp_mom_space}.
We note also that the classical electrodynamics is built in the
four-dimensional Minkowski space.
Formally, this space corresponds to the space of vectors whose three (space)
components are purely imaginary and the fourth (time) component is real.
The rotations in Minkowski space are represented by the Lorentz group.
The multipole expansion technique developed below remains the same also in
Minkowski space.

As an example of the developed formalism, several expansion formulas are
derived for tensor functions depending on the linear vector combination
$\vvr_1+\vvr_2$.
The proposed method can easily be generalised to the case of many-dimensional
spaces.
However, this requires the extensive study of the properties of corresponding
Clebsch-Gordan coefficients which is not the subject of the present paper.

The paper is organised as follows.
In Section~\ref{sec:hypersph-harm-d4} the general properties of
four-dimensional HSH are discussed.
Unlike three-dimensional spherical harmonics where there is only one set of
their indices (and arguments) exist, HSH in four-dimensions can be
parametrised in several ways.
These are analysed in ref.~\cite{aquilanti98prl:_o4_recouplings}.
Below we consider only two sets most important for practical applications.
Keeping in mind the hydrogenic $O(4)$-symmetry, we denote them as
spherical-type and parabolic-type HSH.
In Section~\ref{sec:addit-theor-hypersph} the multipole expansion differential
technique for four-dimensions HSH is developed.
It is based on the rearrangement formula for the exponent 
$\exp (\va \cdot \nabla)$ where $\nabla$ is the gradient operator and $\va$ is
an arbitrary vector in four-dimensional space.
Although the expressions for the Clebsch-Gordan coefficients (CGC) of the
$O(4)$ group are known (see 
e.g. \cite{dolginov59:_clebsch_lorentz_group,biedenharn60jmp:_o4_cgc}),
the definitions used in common literature may differ in phase.
Therefore, below the explicit expressions for CGC are considered.
Analysis shows that for some particular values of indices, CGC can be
written in a compact closed form.
The corresponding expressions are given in
Appendix~\ref{sec:expr-four-dimens}.


\section{Hyperspherical harmonics in four-dimensional space}
\label{sec:hypersph-harm-d4}

In this section the properties of irreducible tensors in four-dimensional
space are analysed.
The Cartesian components of the vector $\vvr$ in four-dimensional space we
denote as $(x,y,z,z_0)$.
We will also use the set of hyperspherical coordinates
$r,\theta_0,\theta,\phi$ which are connected to Cartesian coordinates by means
of identities
\begin{equation}
  \label{eq:hyp-dec-1}
  \begin{split}
    x =& r \sin\theta_0 \sin\theta \cos\phi, \\
    y =& r \sin\theta_0 \sin\theta \sin\phi, \\
    z =& r \sin\theta_0 \cos\theta, \\
    z_0 =& r \cos\theta_0.
  \end{split}
\end{equation}
As is seen $\theta_0$ is the angle between the $z_0$-axis of the coordinate
frame and the vector $\vvr$.
(We note that in Minkowski space the length is defined as
$r^2=z_0^2-x^2-y^2-z^2$ and the replacements $\theta_0 \to i\theta_0$
should be made.)
The inverse relations are
\begin{equation}
  \label{eq:hyp-dec-2}
  \begin{split}
    r^2 =& x^2 + y^2 + z^2 + z^2_0, \\
    \theta_0 =& \arccos \frac{z_0}{r}, \\
    \theta =& \arctan\frac{\sqrt{x^2+y^2}}{z}, \\
    \phi =& \arctan\frac{y}{x}.
  \end{split}
\end{equation}
(The inverse relations in Minkowski space can be derived similarly.)

By definition, an irreducible tensor is a traceless\footnote{That means, the
 contraction of any pair of tensor indices gives zero.}
tensor, symmetric with
respect to the interchange of any pair of its indices \cite{landau}.
The count of different components of the rank-$j$ irreducible tensor in
four-dimensional space gives the number $(j+1)^2$.
This means that the irreducible tensor can be labelled with two indices each
of which runs from $0$ to $j$.
Tensor products $C_{j, \mu, \nu} (\hr)$ will be denoted as
\textit{parabolic-type} set of HSH.
There exists another possibility of choosing the set of tensor indices.
Namely, one can use indices $\lambda$ and $\alpha$ so that $\lambda$ varies
from $0$ to $j$ and $\alpha$ runs from $-\lambda$ to $\lambda$.
The total number of all possible combinations of indices $\lambda$ and
$\alpha$ remains, of course, equal to $(j+1)^2$.
The set of functions $C_{j, \lambda, \alpha} (\hr)$ will be referred to as 
\textit{spherical-type} HSH.

It is easy to see that the irreducible tensor product of $n$ vectors $\vvr$
satisfies the Laplace equation,
\begin{equation}
  \label{eq:4d-lapl-eq}
  \begin{split}
&  \Delta \{ \vvr \}_{j, \mu, \nu} = 0, \\
& \Delta = \frac{\partial^2 }{\partial x^2}
+ \frac{\partial^2 }{\partial y^2} + \frac{\partial^2 }{\partial z^2}
+ \frac{\partial^2 }{\partial z_0^2}.    
  \end{split}
\end{equation}
Indeed, the tensor $\{ \vvr \}_{j, \mu, \nu}$ is a homogeneous polynomial of
the order $j$ with respect to the components $r_{1, \mu, \nu}$.
The action of the scalar Laplace operator $\Delta$ on such a polynomial
decreases its order by two.
However, the result of this action should still be an irreducible tensor of
the same rank $j$ because the tensor equality must contain only tensors of
equal ranks.
Thus, the above statement is proved by contradiction.
We note also that the tensor product
$\{ \vvr \}_{j, \mu, \nu} = r^j \{ \hr \}_{j, \mu, \nu}$
satisfies the boundary condition
\begin{equation}
  \label{eq:bound_cond1}
 \lim_{r \to 0} \{ \vvr \}_{j, \mu, \nu} \to 0,
\end{equation}
i.e. it vanishes at the origin.
Another tensor solution of the four-dimensional Laplace equation which is
divergent at the origin has the form $r^{-j-2} \{ \hr \}_{j, \mu, \nu}$.
Clearly, the tensor product of \textit{unit} vectors $\hr$, 
$\{ \hr \}_{j, \mu, \nu}$ is independent of the hyperradius $r$, and, hence,
it is the function of hyperangles.
Therefore, the tensor product $\{ \hr \}_{j, \mu, \nu}$ coincides with HSH up
to normalisation factor.
(The same is true also for the set of indices $j,\lambda,\alpha$.)

Below, we consider two possible parametrisations for the four-dimensional
HSH.
In sec.~\ref{sec:hyperb-type-spher} the parabolic-type set of HSH is
introduced.
This set is particularly convenient for the calculations since the
corresponding Clebsch-Gordan coefficients have simple form.
The spherical-type HSH are discussed in sec.~\ref{sec:spher-k-harm}.
Such harmonics describe the wave functions of the hydrogen atom in momentum
space labelled with spherical quantum numbers $n,l,m$.
The Clebsch-Gordan coefficients for spherical-type HSH have more complicated
structure comparing to that of the parabolic-type HSH.


\subsection{Parabolic-type spherical harmonics}
\label{sec:hyperb-type-spher}

Now let us introduce the \textit{hyperspherical components} of the vector
$\vvr$ according to the relations
\begin{equation}
  \label{eq:hyp-comp-r}
  \begin{split}
      r_{\pm\halb, \pm\halb} =& \frac{1}{\sqrt2} (z_0 \mp i z), \\
      r_{\pm\halb, \mp\halb} =& - \frac{i}{\sqrt2} (x \mp i y).
  \end{split}
\end{equation}
From these equations it follows that
\begin{equation}
  \label{eq:cc-rmunu}
  r_{\mu, \nu}^* = (-1)^{\mu-\nu} r_{-\mu, -\nu}, \quad \mu,\nu=-\halb,\halb.
\end{equation}

The square length of $\vvr$ in terms of these components has the form
\begin{equation}
  \label{eq:length-1}
  r^2 = \sum_{\mu,\nu=-\halb}^\halb r_{\mu, \nu} r^*_{\mu, \nu}
= \sum_{\mu,\nu} (-1)^{\mu-\nu} r_{\mu, \nu} r_{-\mu, -\nu}
= x^2 + y^2 +z^2 + z_0^2.
\end{equation}
The hyperspherical components of $\vvr$ are connected to the hyperangles as
\begin{equation}
  \label{eq:comp-angles-1}
  \begin{split}
   r_{\pm\halb, \pm\halb} =& \frac{r}{\sqrt2} 
 (\cos\theta_0 \mp i \sin\theta_0 \cos\theta), \\
   r_{\pm\halb, \mp\halb} =& \pm i \frac{r}{\sqrt2} 
 \sin\theta_0 \sin\theta e^{\mp i \phi}.
  \end{split}
\end{equation}
These equation can be re-written in terms of three-dimensional finite rotation
matrices $U^{\halb}_{\mu \nu}$ parametrised by the direction of rotation and
the rotational angle \cite{Varsh},
\begin{equation}
  \label{eq:r-d-1}
  r_{\mu, \nu} = \frac{r}{\sqrt2} U^{1/2}_{\mu, \nu} (2 \theta_0,\theta,\phi).
\end{equation}

We define the parabolic-type (or $H$-harmonics)
spherical harmonics $H_{j, \mu, \nu} (\hr)$ according to the relation
\begin{equation}
  \label{eq:solid-harm}
\{ \hr \}_{j, \mu, \nu} = \frac{1}{2^{j/2}} H_{j, \mu, \nu} (\hr)
= \frac{1}{2^{j/2}} U^{j/2}_{\mu, \nu} (2 \theta_0,\theta,\phi),
\end{equation}
where $U^{j/2}_{\mu, \nu}$ are elements of the three-dimensional finite
rotation matrix.
Written in terms of Euler angles these matrix elements are called Wigner's
$D$-functions.

Let us consider $H$-harmonics for some specific values of its arguments and
indices.
The evaluation of the harmonics $H_{j, \mu, \nu} (\ve_z)$ depending on the
unit vector $\ve_0$ with components $\ve_0=(0,0,0,1)$ can easily be
performed noting that its spherical coordinates are $r=1$,
$\theta_0=\theta=\phi=0$.
For this case we have \cite{Varsh},
\begin{equation}
  \label{eq:c-e-z}
  H_{j, \mu, \nu} (\ve_{z_0}) = U^{j/2}_{\mu, \nu} (0,0,0)
= \delta_{\mu,\nu}.
\end{equation}
Note that $\ve_0$ is directed along the $z_0$-axis of the coordinate frame.

The scalar product of two $H$-harmonics is defined as
\begin{equation}
  \label{eq:sprod-2h}
  (H_j (\ha) \cdot H_j (\hb))
= \sum_{\mu,\nu=-j/2}^{j/2} (-1)^{\mu-\nu} H_{j, \mu, \nu}(\ha)
H_{j, -\mu, -\nu}(\hb) = C^1_j (\ha \cdot \hb),
\end{equation}
where $C^1_j(\ha \cdot \hb)$ is the Gegenbauer polynomial.
The expansion formula for the product of two hyperspherical harmonics
$H_{j, \mu, \nu}$ can be derived based on properties of the $3d$-finite
rotation matrices $U^{j/2}_{\mu, \nu}$,
\begin{equation}
  \label{eq:add-t-1}
  H_{j_1, \mu_1, \nu_1} (\hr)  H_{j_2, \mu_2, \nu_2} (\hr)
= \sum_{j=|j_1-j_2|}^{j_1+j_2} 
H^{j \mu \nu}_{j_1 \mu_1 \nu_1;\, j_2 \mu_2 \nu_2}
H_{j, \mu, \nu}(\hr),
\end{equation}
where $H^{j \mu \nu}_{j_1 \mu_1 \nu_1;\, j_2 \mu_2 \nu_2}$ are the
``parabolic-type'' CGC for the $O(4)$ group.
From the definition (\ref{eq:solid-harm}) of $H$-harmonics as $3d$ finite
rotation matrices it follows that H-type CGC are simply products of two CGC of
the $O(3)$ group
\begin{equation}
  \label{eq:cg-coefs-1}
H^{j, \mu, \nu}_{j_1 \mu_1 \nu_1;\, j_2 \mu_2 \nu_2}
= C^{(j/2)\mu}_{(j_1/2)\mu_1\, (j_2/2) \nu_2} 
C^{(j/2) \nu}_{(j_1/2) \nu_1\, (j_2/2)\nu_2}.
\end{equation}
This expression for CGC directly follows from the definition of spherical
harmonics as the elements of the finite rotation matrix in $3d$-space.
We will not discuss further properties of coefficients 
$H^{j \alpha \alpha'}_{l \mu \mu';\, l'\nu\nu'}$, such as orthogonality
\textit{etc}, since they can be easily deduced from the properties of
conventional CGC in $3d$-space.

From the expression \eqref{eq:cg-coefs-1} for CGC one can deduce the triangle
rule for the momenta $j_1,j_2,j$.
Namely, the momentum $j$ can take the values
$|j_1-j_2|, |j_1-j_2|+2, \ldots, j_1+j_2$.
Thus, the sum $j_1+j_2+j_3$ is always an even number.
In particular, for $j_1=j_2$ we have that $j=0,2,4\ldots, (2j_1)$.

The tensor product of two spherical harmonics we will denote as
$\{H_{j_1} (\ha) \otimes H_{j_2} (\hb) \}_j$ and it is defined by the equation
\begin{equation}
  \label{eq:4d-bh-1}
  \{H_{j_1} (\ha) \otimes H_{j_2} (\hb) \}_{j, \mu, \nu}
= \sum_{\mu_1,\nu_1=-j_1/2}^{j_1/2} \sum_{\mu_2,\nu_2=-j_2/2}^{j_2/2} 
H^{j, \mu, \nu}_{j_1 \mu_1 \nu_1;\, j_2 \mu_2 \nu_2}
H_{j_1, \mu_1, \nu_1}(\ha) H_{j_2, \mu_2, \nu_2}(\hb).
\end{equation}
Below, we will label the tensor products and spherical harmonics only with
their ranks unless one needs to derive explicit expressions.
The bipolar harmonics of the zero rank are connected with the scalar product
of spherical harmonics as
\begin{equation}
  \label{eq:bh0}
  \{H_l (\ha) \otimes H_l (\hb) \}_{0}
= \frac{1}{l+1} (H_l (\ha) \cdot H_l (\hb)).
\end{equation}
For two equal vectors $\ha=\hb$ this equation becomes
\begin{equation}
  \label{eq:bh-00}
  \{H_l (\ha) \otimes H_l (\ha) \}_{0}
= \frac{1}{l+1} (H_l (\ha) \cdot H_l (\ha)) = 1.  
\end{equation}


\subsection{Spherical-type spherical harmonics}
\label{sec:spher-k-harm}

The explicit expression for the spherical-type HSH can be derived using the 
parametrisation of the $3d$ finite rotation matrix in terms of the rotation
angle and the rotation axis,
\begin{equation}
  \label{eq:d-u}
  U^l_{\mu, \nu}(2 \theta_0, \theta, \phi)
= \sum_{\lambda=0}^{2l}
 (-i)^\lambda \frac{2\lambda+1}{2l+1} \,
  C^{l\nu}_{l\mu \, \lambda \alpha} \chi^l_\lambda (2\theta_0) 
 C_{\lambda \alpha} (\theta,\phi),
\end{equation}
where $C_{\lambda \alpha}$ are modified spherical harmonics connected to the
usual $3d$-spherical harmonics by 
$C_{\lambda \alpha} = \sqrt{4\pi/(2\lambda+1)} Y_{\lambda \alpha}$.
 The generalised characters of the $O(3)$ rotation group $\chi^l_\lambda$
can be expressed in terms of Gegenbauer polynomials \cite{Varsh},
\begin{equation}
  \label{eq:chi-geg}
  \chi^l_\lambda (2 \theta_0)
= (2 \lambda)!!\, \sqrt{2l+1} \, \sqrt\frac{(2l-\lambda)!}{(2l+\lambda+1)!}
(\sin\theta_0)^\lambda\, C^{\lambda+1}_{2l-\lambda}(\cos\theta_0).
\end{equation}

We introduce the set of spherical-type ($C$-harmonics) HSH 
$C_{j,\lambda, \mu}$ by connecting them to $H$-harmonics as
\begin{equation}
  \label{eq:c-1}
  C_{j, \lambda, \mu} (\hr)
= \sqrt\frac{2 \lambda +1}{j+1} \sum_{\mu,\nu=-j/2}^{j/2}
 C^{(j/2) \mu}_{(j/2) \nu\, \lambda \alpha} H_{j, \mu,\nu} (\hr),
\end{equation}
where $\lambda=0,1,\ldots j$, 
and $\alpha=-\lambda,-\lambda+1,\ldots \lambda$.
The inverse relation can be derived using the orthogonality of $3d$ CGC
\begin{equation}
  \label{eq:h-c}
  H_{j, \mu, \nu} (\hr) = 
\sum_{\lambda=0}^{j} 
 \sqrt\frac{ 2 \lambda+1}{j+1} \,
 C^{(j/2) \mu}_{(j/2) \nu\, \lambda \alpha} 
C_{j, \lambda, \alpha} (\hr).
\end{equation}
The explicit expression for $C$-harmonics can be obtained from (\ref{eq:d-u})
and it is
\begin{equation}
  \label{eq:def-h}
  C_{j, \lambda, \alpha} (\hr)
= (-i)^\lambda \sqrt\frac{ 2 \lambda+1}{j+1} 
\chi^{j/2}_\lambda (2 \theta_0) C_{\lambda \alpha } (\theta,\phi).
\end{equation}
From this equation follows the formula for complex conjugated harmonics
\begin{equation}
  \label{eq:c-2}
  C^*_{j, \lambda, \alpha} (\hr) = (-1)^{\lambda+\alpha} 
C_{j, \lambda, -\alpha} (\hr).
\end{equation}
The connection of $C$-harmonics with tensor products of vectors is the
same as for $H$-harmonics which is given by eq.~(\ref{eq:solid-harm}).
The scalar product of two $C$-harmonics is defined similarly to the above
eq.~(\ref{eq:sprod-2h}),
\begin{equation}
  \label{eq:c-3}
  (C_j (\ha) \cdot C_j (\hb))
= \sum_{\lambda, \alpha} C^*_{j, \lambda, \alpha}(\ha) 
C_{j, \lambda, \alpha}(\hb)
= \sum_{\lambda, \alpha}  (-1)^{\lambda+\alpha}
C_{j, \lambda, \alpha}(\ha) C_{j, \lambda, -\alpha}(\hb)
= C_j^1 (\ha \cdot \hb).
\end{equation}
We note also the expression for $C_{j}(\ve_0)$ which can be derived from
eq.~(\ref{eq:c-1}),
\begin{equation}
  \label{eq:c-e0}
  C_{j, \lambda, \alpha} (\ve_0) = 
 (-i)^\lambda \sqrt\frac{ 2 \lambda+1}{j+1} 
\chi^{j/2}_\lambda (0) C_{\lambda \alpha } (0,0)
= \sqrt{j+1}\, \delta_{\lambda,0} \delta_{\alpha,0},
\end{equation}
where we have used the property of the generalised characters \cite{Varsh}
\begin{displaymath}
  \chi^{j/2}_\lambda (0)= (j+1) \,\delta_{\lambda,0}.
\end{displaymath}
Now we have to establish the expression for CGC for the set of $C$-harmonics.
We define those coefficients by considering the expansion of the product of
two $C$-harmonics 
\begin{equation}
  \label{eq:c-4}
C_{j_1, \lambda_1, \alpha_1} (\hr) C_{j_2, \lambda_2, \alpha_2} (\hr)
= \sum_{j \lambda \alpha} 
C^{j \lambda \alpha}_{j_1 \lambda_1 \alpha_1;\, j_2 \lambda_2 \alpha_2}
C_{j, \lambda, \alpha} (\hr).
\end{equation}
CGC in this equation can be calculated by expressing each $C$-harmonic in
terms of $H$-harmonics according to the decomposition (\ref{eq:c-1}).
At this stage the CGC for $H$-harmonics will occur.
However, these are already known, see eq.~(\ref{eq:cg-coefs-1}).
After somewhat lengthy manipulations with $3d$ CGC one arrives at the
expression
\begin{equation}
  \label{eq:clebsh-c}
C^{j \lambda \alpha}_{j_1 \lambda_1 \alpha_1;\, j_2 \lambda_2 \alpha_2}
= (j+1) \sqrt{(2\lambda_1+1)(2\lambda_2+1)}
C^{\lambda \alpha}_{\lambda_1 \alpha_1\, \lambda_2 \alpha_2}
\left\{
  \begin{array}{ccc}
    \frac{j_1}{2} & \frac{j_2}{2} & \frac{j}{2} \\
    \frac{j_1}{2} & \frac{j_2}{2} & \frac{j}{2} \\
    \lambda_1 & \lambda_2 & \lambda
  \end{array}
\right\},
\end{equation}
where the table in curly brackets is the $9j$-coefficient of $3d$-space.
The $9j$-coefficient with two identical rows is non-zero only if the sum of its
indices is an even number \cite{Varsh}.
Note, that the sum $j_1+j_2+j$ is always an even number.
Hence, the $C$-type CGC is non-zero only if the combination 
$\lambda_1+\lambda_2+\lambda$ is an even number.
Thus, the parameter $\lambda$ can be 
$|\lambda_1-\lambda_2|, |\lambda_1-\lambda_2|+2, \ldots, \lambda_1+\lambda_2$.

The bipolar harmonics of $C$-type are defined similarly to the $H$-case
(\ref{eq:4d-bh-1}),
\begin{equation}
  \label{eq:bh-c}
   \{C_{j_1} (\ha) \otimes C_{j_2} (\hb) \}_{j, \lambda, \alpha}
= \sum_{\lambda_1,\lambda_2,\alpha_1,\alpha_2} 
C^{j \lambda \alpha}_{j_1 \lambda_1 \alpha_1;\, j_2 \lambda_2 \alpha_2}
C_{j_1, \lambda_1 \alpha_1}(\ha) C_{j_2, \lambda_2, \alpha_2}(\hb). 
\end{equation}
The expressions (\ref{eq:bh0}), (\ref{eq:bh-00}) for the scalar product of
$H$-harmonics remain valid after the replacement $H \leftrightarrow C$.

We have to analyse the properties of the $C$-type CGC.
Clearly, they follow from properties of the $3d$ CGC and $9j$-symbols.
For example, the orthogonality relation of $4d$ CGC has the form
\begin{equation}
  \label{eq:cg-ort}
  \sum_{\lambda_1, \lambda_2, \alpha_1, \alpha_2}
C^{j \lambda \alpha}_{j_1 \lambda_1 \alpha_1;\, j_2 \lambda_2 \alpha_2}
C^{j' \lambda' \alpha'}_{j_1 \lambda_1 \alpha_1;\, j_2 \lambda_2 \alpha_2}
=\delta_{j,j'} \delta_{\lambda,\lambda'} \delta_{\alpha,\alpha'}.
\end{equation}
The exchange symmetry of $4d$ CGC is
\begin{equation}
  \label{eq:cg-1}
 C^{j \lambda \alpha}_{j_1 \lambda_1 \alpha_1;\, j_2 \lambda_2 \alpha_2}
= (-1)^{j_1+j_2+j}
C^{j \lambda \alpha}_{j_2 \lambda_2 \alpha_2; \, j_1 \lambda_1 \alpha_1}
\end{equation}
We note that the above identities hold for both, $C$- and $H$-types of CGC.
We present also one more symmetry relation valid for $C$-type CGC
\begin{equation}
  \label{eq:cg-2}
C^{j \lambda \alpha}_{j_1 \lambda_1 \alpha_1; \, j_2 \lambda_2 \alpha_2}
= (-1)^{j_1+j_2+j+\lambda_2 +\alpha_2}
\frac{j+1}{j_1+1}
C^{j_1 \lambda_1 \alpha_1}_{j \lambda \alpha;\, j_2 \lambda_2 -\alpha_2}.
\end{equation}
At some particular values of its parameters the $C$-type CGC may be written in
closed form.
For example, this happens at $j_1=0$ where
\begin{equation}
  \label{eq:cg-00}
C^{j \lambda \alpha}_{0 0 0;\, j_2 \lambda_2 \alpha_2}
= \delta_{j,j_2} \delta_{\lambda,\lambda_2} \delta_{\alpha,\alpha_2}.
\end{equation}
For zero projections, $C$-type CGC also evaluates to a closed form,
\begin{equation}
  \label{eq:cgc-zero-proj}
 C^{j 00}_{j_1 00;\, j_2 00} = \sqrt\frac{j+1}{(j_1+1)\,(j_2+1)}.
\end{equation}
More explicit expressions for $C$-type CGC are given in
Appendix~\ref{sec:expr-four-dimens}.

We have to consider also the re-coupling coefficients in $4d$-space.
These re-coupling coefficients can depend only upon the ranks of tensors but
not on their projection indices.
Therefore, all tensor identities are covariant, i.e. they are equally valid
for $H$- and $C$-type tensor components.
As an example, we consider the re-coupling relation
\begin{multline}
  \label{eq:recoupl}
  \{ \{P_a \otimes Q_b \}_c \otimes \{R_d \otimes S_e \}_f \}_k
= \sum_{gh} (c+1)(f+1)(g+1)(h+1) \\
\times
  \{ \{P_a \otimes R_d \}_g \otimes \{Q_b \otimes S_e \}_h \}_k
\left[
  \begin{array}{ccc}
   a & b & c\\
   d & e & f\\
   g & h & k
  \end{array}
\right],
\end{multline}
where the table in square brackets is the $9j$-symbol in $4d$-space.
Its calculation is more simple when using $H$-type CGC.
Omitting the details of derivations we present only the result,
\begin{equation}
  \label{eq:9j-04}
\left[
  \begin{array}{ccc}
   a & b & c\\
   d & e & f\\
   g & h & k
  \end{array}
\right]=
  \left\{
  \begin{array}{ccc}
   \frac{a}{2} & \frac{b}{2} & \frac{c}{2}\\
   \frac{d}{2} & \frac{e}{2} & \frac{f}{2}\\
   \frac{g}{2} & \frac{h}{2} & \frac{k}{2}
  \end{array}
\right\}^2
\end{equation}
As is seen, the irreducible tensor in $4d$-space may be considered as
``double-tensors'' in $3d$-space.

Finally, we note that HSH are orthogonal,
\begin{multline}
  \label{eq:c-orthog-1}
  \int_S C_{j, \lambda, \alpha} (\hr)
C^*_{j', \lambda', \alpha'} (\hr) \, d \Omega
= \int_0^\pi \sin^2 \theta_0 \, d \theta_0 
\int_0^\pi \sin \theta \,d\theta \int_0^{2\pi} d\phi\,
C_{j, \lambda, \alpha} (\theta_0,\theta,\phi)
C^*_{j', \lambda', \alpha'} (\theta_0,\theta,\phi) \\
= \frac{2 \pi^2}{j+1} \,
\delta_{j,j'} \delta_{\lambda,\lambda'} \delta_{\alpha,\alpha'},
\end{multline}
where the integration is performed over the four-dimensional hypersphere $S$.
Often, HSH normalised to unity are used
\begin{equation}
  \label{eq:def-y}
  Y_{j,\lambda,\alpha} (\hr) = \frac{(-1)^{j+\lambda}}{\pi} \sqrt\frac{j+1}{2}
   C_{j,\lambda,\alpha} (\hr).
\end{equation}
The orthogonality identity holds also for $H$-type spherical harmonics,
\begin{equation}
  \label{eq:c-orthog-2}
  \int_S H_{j, \mu, \nu} (\hr)
H^*_{j', \mu', \nu'} (\hr) \, d \Omega  
= \frac{2 \pi^2}{j+1} \, \delta_{j,j'} \delta_{\mu,\mu'} \delta_{\nu,\nu'}.
\end{equation}


\section{The multipole expansions in hyperspherical harmonics}
\label{sec:addit-theor-hypersph}

In this section the differential formalism for the derivation of multipole
expansions in four-dimensional space is developed.
The general formulas are derived in Sec.~\ref{sec:diff-mult-expans}.
In Sec.~\ref{sec:mult-expans-power} we consider the multipole expansions of
the function $| \va + \vvr|^n C_j (\hb)$ where $\hb$ is the unit vector
directed along the vector sum $(\va+\vvr)$ .


\subsection{The differential multipole expansion formula}
\label{sec:diff-mult-expans}

We start with the conventional Taylor expansion formula in
many-dimensional space:
\begin{equation}
  \label{eq:taylor-1}
  f(\va+\vvr) = \sum_{l=0}^\infty \frac{1}{l!} (\va \cdot \nabla)^l f(\vvr).
\end{equation}
It can be re-written in the simple symbolic form,
\begin{equation}
  \label{eq:taylor-symb}
  f(\va+\vvr) = e^{(\va \cdot \nabla)} f(\vvr).  
\end{equation}
We have to present this formula in a way suitable for the derivation of
multipole expansions of the function $f(\vvr)$.
This can be done similarly to the three-dimensional case considered in
\cite{manakov02:_multip}.
Namely, one has to use the multipole expansion for the operator exponent
$\exp{(\va \cdot \nabla)}$.
Before doing so, we write the multipole expansion of the exponential
scalar product in four-dimensional space \cite{avery85jmp:_hyp_spher}, 
\begin{equation}
  \label{eq:wen-avery}
 e^{\va \cdot \vvr}
= \sum_{l=0}^\infty \frac{i^{l+1}}{a r} 2 (l+1) J_{l+1}(-i ar) 
C^1_l (\ha \cdot \hr).
\end{equation}
The  Bessel function $J_{l+1}(iar)$ can be written in terms of a
hypergeometric function \cite{Bateman-II} as
\begin{equation}
  \label{eq:bessel-hypg}
  J_{l+1} (-ix) = \frac{(-ix)^{l+1}}{2^{l+1} (l+1)!}\, {}_0F_1 (l+2; x^2/4).
\end{equation}
This formula allows one to re-write the decomposition (\ref{eq:wen-avery}) in
the form
\begin{equation}
  \label{eq:exp-hgeom}
  e^{\va \cdot \vvr} 
= \sum_{l=0}^\infty \frac{ a^l r^l }{2^l l!}
\, {}_0F_1 (l+2; a^2 r^2/4) C^1_l (\ha \cdot \hr)
= \sum_{l=0}^\infty \frac{ a^l }{2^{l/2} l!}
\, {}_0F_1 (l+2; a^2 r^2/4) 
(C_l (\ha) \cdot \{ \vvr \}_l).
\end{equation}
 This equation can be used for the derivation of the multipole decomposition
for the scalar products $(\va \cdot \vvr)^n$.
Namely, one has to expand the exponent $\exp{(\va \cdot \vvr)}$ and the
hypergeometric function ${}_0F_1 (l+2; a^2 r^2/4)$ into the power series and
compare the coefficients at equal powers of $ar$.
This leads to the formula
\begin{multline}
  \label{eq:ar-geg-1}
  (\va \cdot \vvr)^n =  n!
  \left(
    \frac{ar}{2}
  \right)^n
\sum_{k=0}^{[n/2]} \frac{n-2k+1}{k! (n-k+1)!} C^1_{n-2k}(\ha \cdot \hr) \\
= n! (ar)^n \sum_{l=n,n-2,\ldots} \frac{2 (l+1)}{(n-l)!! (n+l+2)!!} 
(C_l (\ha) \cdot C_l(\hr)).
\end{multline}
For small values of $n$ this formula has been checked explicitly.

Replacing in eq.~(\ref{eq:exp-hgeom}) $\vvr$ with $\nabla$ one arrives at the
following equation for the Taylor expansion (\ref{eq:taylor-symb}),
\begin{equation}
  \label{eq:taylor-2}
 f(\va+\vvr) =  \sum_{l=0}^\infty \frac{ a^l }{2^{l/2} l!}
\, {}_0F_1 (l+2; a^2 \Delta/4) 
(C_l (\ha) \cdot \{ \nabla \}_l) f(\vvr).
\end{equation}
This is the key equation for the derivation of multipole expansions.

For further consideration it is necessary to calculate the action of the
Laplace operator on HSH,
\begin{equation}
 \label{eq:lapl-2}
 \Delta C_l (\hr) =
 \left(
  \frac{1}{r^3} \frac{\partial}{\partial r} r^3
\frac{\partial}{\partial r} 
+ \frac{\Delta_\Omega}{r^2}
 \right) C_l (\hr) = 
\frac{\Delta_\Omega}{r^2} C_l(\hr) 
= - \frac{l (l+2)}{r^2} C_l (\hr).
\end{equation}
Hereafter, we will label HSH only with their ranks and, for the sake of
shortness, we will omit all projection indices.

Using the above formula one can prove the identity
\begin{equation}
  \label{eq:lapl-act-1}
  \Delta f(r) C_l (\hr) = 
C_l (\hr)
\left(
  \frac{1}{r^3} \frac{\partial}{\partial r} r^3
\frac{\partial}{\partial r} 
- \frac{l (l+2)}{r^2}  
\right) f(r),
\end{equation}
where $f(r)$ is an arbitrary function depending on $r = | \vvr|$.
This equation can be presented in two different compact forms
\begin{equation}
  \label{eq:lapl-fc}
  \Delta f(r) C_l (\hr) = C_l (\hr)\,
\frac{1}{r^{l+3}} \frac{\partial}{\partial r} r^{2l+3}
\frac{\partial}{\partial r} \frac{1}{r^l} f(r)
= C_l (\hr) r^{l-1}
\frac{\partial}{\partial r} \frac{1}{r^{2l+1}}
 \frac{\partial}{\partial r} r^{l+2} f(r).
\end{equation}
In the important particular case of $f(r)=r^n$ these identities lead to the
formula
\begin{equation}
  \label{eq:rn-1}
  \Delta^k r^n C_l(\hr)=  C_l(\hr) \, r^{n-2k}
\frac{(n-l)!!}{(n-l-2k)!!} \frac{(l+n+2)!!}{(l+n+2-2k)!!}.
\end{equation}
One can re-write this equation in a form free of factorials
\begin{equation}
  \label{eq:rn-2}
  \Delta^k r^n C_l(\hr) =  C_l(\hr) r^{n-2k}
2^{2k} \left(\frac{-2 -l-n}{2} \right)_k \left(\frac{l-n}{2}\right)_k,
\end{equation}
where $(a)_k=\Gamma(a+k)/\Gamma(a)$ is the Pochhammer symbol.
As is seen, this equation is also valid in the case of non-integer values of
$n$.


\subsection{Multipole expansions of translated hyperspherical harmonics}
\label{sec:mult-expans-power}

Below we calculate the multipole expansion of the function 
$f(\va+\vvr)=|\va + \vvr|^n C_j (\hb)$, where
$\hb=(\va+\vvr)/|\va+\vvr|$.
According to eq.~(\ref{eq:taylor-2}), in order to calculate the multipole
expansion of this function, one has to calculate the operator action
\begin{displaymath}
 (C_l (\ha) \cdot \{ \nabla \}_l) f(\vvr)
= (C_l (\ha) \cdot \{ \nabla \}_l) r^n C_j (\hr).
\end{displaymath}
From the general symmetry arguments it is clear that this action can be
presented as a combination of bipolar harmonics
\begin{equation}
  \label{eq:partf-1}
  (C_l (\ha) \cdot \{ \nabla \}_l) r^n C_j (\hr)
= r^{n-l} \sum_{l'} A_{ll'} \{C_l (\ha) \otimes C_{l'} (\hr) \}_j,
\end{equation}
where the coefficients $A_{ll'}$ are numbers which depend apart of $l,l'$
also on $j$ and $n$ but are independent on vectors $\va$ and $\vvr$.

The independence of the coefficients $A_{ll'}$ on vectors $\va$ and $\vvr$
allows one to simplify their calculation.
We will calculate $A_{ll'}$ for vectors $\va$ having zero length: 
$(\va \cdot \va)=0$.
Noting this fact and the expression (\ref{eq:c-3}) for the scalar product of
$C$-harmonics we arrive at the identity
\begin{equation}
  \label{eq:a-1}
  (C_l (\ha) \cdot \{ \nabla \}_l) = 2^{l/2} (\va \cdot \nabla)^l.
\end{equation}
Now one has to evaluate the action of the operator in the rhs of this
equation on the product $r^nC_j(\hr)$. 
This can be done using the chain differentiation rule
\begin{equation}
  \label{eq:a-2}
  (\va \cdot \nabla)^l r^n C_j (\hr)
= 2^{j/2} (\va \cdot \nabla)^l r^{n-j} \{ \vvr \}_j
= 2^{j/2}
\sum_{k=0}^l \binom{l}{k} \left[ (\va \cdot \nabla)^k r^{n-j} \right]
(\va \cdot \nabla)^{l-k} \{ \vvr \}_j,
\end{equation}
where $\binom{l}{k}= l!/(k!(l-k)!)$ is the binomial coefficient and
$\nabla$-operators in square brackets do not act on the outer terms.
The term $(\va \cdot \nabla)^{l-k} \{ \vvr \}_j$ can be evaluated using the
vector differentiation technique described in \cite{Man-96} so that
\begin{equation}
  \label{eq:a-3}
  (\va \cdot \nabla)^{l-k} \{ \vvr \}_j
= \frac{j!}{(j-l+k)!} \{\{\va\}_{l-k} \otimes \{\vvr\}_{j-l+k}\}_j.
\end{equation}
The calculation of operator action in square brackets in (\ref{eq:a-2})
simplifies because $(\va \cdot \va)=0$.
After some simple analysis one obtains
\begin{multline}
  \label{eq:a-4}
  (\va \cdot \nabla)^k r^{n-j}
= (n-j) (n-j -2 ) \cdots (n-j-2k+2) (\va \cdot \vvr)^k r^{n-j-2k} \\
=(-2)^k \left(-\frac{n-j}{2} \right)_k (\va \cdot \vvr)^k r^{n-j-2k}.
\end{multline}

Now we have to re-write the combination 
$(\va \cdot \vvr)^k \{\{ \va \}_{l-k} \otimes \{\vvr\}_{j-l+k}\}_j$ 
in terms of HSH.
This can be achieved using the tensor re-coupling rules.
To illustrate this we note that the above construction can be presented in the
form
\begin{equation}
  \label{eq:a-5}
 (\va \cdot \vvr)^k \{\{ \va \}_{l-k} \otimes \{\vvr\}_{j-l+k}\}_j
= r^{j-l+2k} (\{ \va \}_k \cdot \{ \hr \}_k) \,
\{\{\va\}_{l-k} \otimes \hr\}_{j-l+k}\}_j.
\end{equation}
The re-coupling of the tensor products in rhs of this equation yields
\begin{multline}
  \label{eq:a-6}
  (\{ \va \}_k \cdot \{ \hr \}_k) \,
\{\{\va\}_{l-k} \otimes \{\hr\}_{j-l+k}\}_j
= 2^{(l-j)/2-k} \sum_{l'} (k+1)(j+1) (l+1) (l'+1) \\
\times
\left[
  \begin{array}{ccc}
    k & k & 0 \\
    l-k & j-l+k & j \\
    l & l' & j
  \end{array}
\right]
\{ \{\ha \}_l \otimes C_{l'}(\hr) \}_j,
\end{multline}
where we have used the auxiliary identity
\begin{displaymath}
\{\{\va \}_k \otimes \{\va\}_{l-k} \}_q
= \{ \va\}_l \delta_{lq},
\end{displaymath}
which follows from the fact that $(\va \cdot \va)=0$.
The summation index $l'$ in eq.~(\ref{eq:a-6}) takes the values of
$| j-l|, | j-l| +2, \ldots, j-l+2k$.
Thus, the combination $l+l'+j$ is always an even number.

The four-dimensional re-coupling coefficient in eq.~(\ref{eq:a-6}) is
connected with three-dimensional $9j$-symbols by means of
eq.~(\ref{eq:9j-04}).
The resulting $9j$-coefficient can be evaluated in closed form \cite{Varsh}, so
that the four-dimensional re-coupling coefficient becomes
\begin{multline}
  \label{eq:4d9j-2}
\left[
  \begin{array}{ccc}
    k & k & 0 \\
    l-k & j-l+k & j \\
    l & l' & j
  \end{array}
\right]
= (-1)^{j+l+l'}
\frac{k!\,(j-l+k)!}{ (l+1)!\, (j+1)!\, (k+1)\, (j+1)} \\
\times
\frac{\Gamma \left( \frac{j+l+l'}{2} + 2 \right)
\Gamma \left( \frac{j+l-l'}{2} + 1 \right)}%
{\Gamma \left( \frac{j-l-l'}{2} + k+1 \right)
 \Gamma \left( \frac{j-l+l'}{2} + k+2 \right)}.
\end{multline}
The next step is to insert this equation into eq.~(\ref{eq:a-6}) and
substitute the result together with eq.~(\ref{eq:a-4}) into
eq.~(\ref{eq:a-2}).
This leads to the identity
\begin{multline}
  \label{eq:sum-k-1}
  (\va \cdot \nabla)^l r^n C_j (\hr)
= r^{n-l} 2^{l/2} \sum_{l'} \{\{ \va \}_l \otimes C_{l'} (\hr)\}_j
 (-1)^{j+l+l'} \frac{l'+1}{j+1}  \\
\times
  \sum_{k=0}^l \frac{(-1)^k}{(l-k)!}
  \left(
    \frac{j-n}{2}
  \right)_k
\frac{\Gamma \left( \frac{j+l+l'}{2} + 2 \right)
\Gamma \left( \frac{j+l-l'}{2} + 1 \right)}%
{\Gamma \left( \frac{j-l-l'}{2} + k+1 \right)
 \Gamma \left( \frac{j-l+l'}{2} + k+2 \right)}.
\end{multline}
Here, the summation over $k$ can be performed analytically.
Noting also the identity (\ref{eq:a-1}) the above equation evaluates to
\begin{multline}
 \label{eq:action-1}
  (C_l(\ha) \cdot \{\nabla\}_l) r^n C_j (\hr)
 = r^{n-l} 2^{l/2} \sum_{l'} (-1)^l
 \{ C_l(\ha) \otimes C_{l'} (\hr)\}_j 
\frac{l'+1}{j+1}  \\
\times
  \left(
   \frac{-2-j-n}{2}
  \right)_{(j+l-l')/2}
  \left(
    \frac{j-n}{2}
  \right)_{(l+l'-j)/2},
\end{multline}
where we have used the fact that $(-1)^{j+l+l'}=1$.
It is important to note that the derived equation is valid for arbitrary
vectors $\va$ (i.e. not only for zero-length vectors).

According to eq.~(\ref{eq:taylor-2}), we have to act on
eq.~(\ref{eq:action-1}) with the operator ${}_0F_1 (l+2; a^2 \Delta/4)$.
Thus, one has to calculate the operator construction
\begin{equation}
  \label{eq:act-1}
  {}_0F_1 (l+2; a^2 \Delta/4) r^{n-l} C_{l'} (\hr)
= \sum_{k=0}^\infty \frac{ a^{2k}}{2^{2k} k! (l+2)_k} \Delta^k
r^{n-l} C_{l'}(\hr).
\end{equation}
Equation (\ref{eq:rn-2}) allows one to calculate the action of Laplace
operators $\Delta^k$ on the product $r^{n-l}C_{l'}(\hr)$.
The resulting series leads to the Gauss hypergeometric function,
\begin{multline}
 \label{eq:act-2}
C_{l'}(\hr) \sum_{k=0}^\infty \frac{ r^{n-l-2k} a^{2k}}{ k! (l+2)_k}
 \left(\frac{-2 +l-l'-n}{2} \right)_k 
\left(\frac{l+l'-n}{2}\right)_k \\
= r^{n-l} C_{l'}(\hr) 
{}_2F_1 \left(\frac{-2 +l-l'-n}{2}, \frac{l+l'-n}{2}; l+2; 
 \frac{a^2}{r^2}\right).
\end{multline}
For the  sake of simpler presentation it is convenient to replace the
notations by $\va \to \vvr_1$ and $\vvr \to \vvr_2$.
Now we can write the explicit expression for the multipole expansion of the
product $r^n C_j (\hr)$, where $\vvr=\vvr_1 + \vvr_2$,
\begin{equation}
  \label{eq:mult-exp-1}
  \begin{split}
   r^n C_j (\hr) 
=& \sum_{l,l'=0}^\infty B^{(nj)}_{ll'} (r_1,r_2) \;
\{ C_{l}(\hr_1) \otimes C_{l'} (\hr_2) \}_j, \\
  B^{(nj)}_{ll'} (r_1,r_2) =& 
 r_2^n \left(- \frac{r_1}{r_2} \right)^l
\frac{l'+1}{l!\, (j+1)}
  \left(
  \frac{-2-j-n}{2}
  \right)_{(j+l-l')/2}
  \left(
    \frac{j-n}{2}
  \right)_{(l+l'-j)/2}  \\
& \times 
 {}_2F_1 \left(\frac{-2 +l-l'-n}{2}, \frac{l+l'-n}{2}; l+2; 
 \frac{r_1^2}{r_2^2}\right),
  \end{split}
\end{equation}
where the summations over $l,l'$ are performed over all values at
which $j+l-l' = 0, 2, 4,\ldots$ and $l+l' \ge j$.
Thus, there is only one infinite summation in the above formula.

Since the combination $(l+l'-j)/2$ is a positive integer number,
at negative integer values of $(j-n)/2$ the above multipole series are, in
fact, finite sums.
Indeed, for $(n-j)=2,4,\ldots$, the second Pochhammer symbols in
the expression for $B^{(nj)}_{ll'}$ vanishes at all values of $l,l'$ except
those with $(l+l'-j) \le |j-n|$.

We note also that for $r_1=i$, $r_2=1$ and $j=0$, $n=2N$, where $N$ can be
arbitrary integer, eq.~(\ref{eq:mult-exp-1}) reduces to
\begin{displaymath}
  (2 i\, (\hr_1 \cdot \hr_2))^N = \sum_{l=N,N-2,\ldots} 
     B^{(2N, 0)}_{ll} (i,1)
 \{ C_l (\hr_1) \otimes C_l (\hr_2) \}_0.
\end{displaymath}
Thus, it is the multipole expansion for the powers of a scalar product which
has been derived above in a different way, see eq.~(\ref{eq:ar-geg-1}).

For finite multipole decompositions the question of convergence does not
occur.
The infinite multipole series is convergent only when $r_1 < r_2$.
In the opposite case (i.e. at $r_1 > r_2$) the replacement 
$\vvr_1 \leftrightarrow \vvr_2$ recovers the convergence of the corresponding
series.

Below we present several explicit examples of the multipole expansion
(\ref{eq:mult-exp-1}).
There are two situations when they have particularly simple form.
Namely, at $n=j$ or $n=-j-2$ the product $r^n C_j (\hr)$ satisfies the Laplace
equation
\begin{displaymath}
  \Delta r^j C_j (\hr) = \Delta \frac{1}{r^{j+2}} C_j (\hr) = 0.
\end{displaymath}
As a consequence, the Gauss hypergeometric function in (\ref{eq:mult-exp-1})
is equal to unity.
This is clearly seen also from eq.~(\ref{eq:act-1}).
The corresponding multipole expansions are
\begin{eqnarray}
  \label{eq:mults-cases-a}
  r^j C_j (\hr) &=& \sum_{l=0}^j  \binom{j}{l} r_1^l r_2^{j-l}
  \{ C_{l}(\hr_1) \otimes C_{j-l} (\hr_2) \}_j, \\
 \label{eq:mults-cases-b}
 \frac{1}{r^{j+2}} C_j(\hr) &=& 
\sum_{l=0}^\infty (-1)^l  \frac{r_1^l}{r_2^{j+l+2}}
  \binom{j+l+1}{l}
 \{ C_{l}(\hr_1) \otimes C_{j+l} (\hr_2) \}_j,
\end{eqnarray}
where $\vvr=\vvr_1+\vvr_2$.

Let us consider one more example of the multipole expansion
eq.~(\ref{eq:mult-exp-1}).
At $j=0$ we have to expand the scalar function $|\vvr_1 + \vvr_2|^n$.
In this case $l=l'$, and noting the identity
\begin{equation}
  \label{eq:sp-gegenb}
  \{ C_l (\hr_1) \otimes C_l (\hr_2) \}_0
= \frac{1}{l+1} (C_l (\hr_1) \cdot C_l (\hr_2))
\end{equation}
we can write the multipole decomposition of the function
$|\vvr_1 + \vvr_2|^n$ as
\begin{multline}
  \label{eq:mult-j-zero}
  |\vvr_1 + \vvr_2|^n
= r_2^n \sum_{l=0}^\infty
\left(
  - \frac{r_1}{r_2}
\right)^l
\frac{1}{l!}
\left(
  - \frac{n}{2}
\right)_l
 {}_2F_1 \left(-1 - \frac{n}{2}, l-\frac{n}{2}; l+2; 
 \frac{r_1^2}{r_2^2}\right)\\
\times (C_l (\hr_1) \cdot C_l (\hr_2)).
\end{multline}
We note that for $n$ being an even integer number, the Gauss hypergeometric
function reduces to the associated Legendre polynomial.
As is seen, the derived multipole series are convergent only for $r_1 < r_2$.
In the opposite case $r_1 > r_2$, the replacement
$\vvr_1 \leftrightarrow \vvr_2$ should be made in the above formula.



\section{Conclusion}
\label{sec:conclusion}

The main results of this paper are the multipole expansions given by
eqs.~(\ref{eq:ar-geg-1}), (\ref{eq:mult-exp-1}), (\ref{eq:mults-cases-a}),
 (\ref{eq:mults-cases-b}) and (\ref{eq:mult-j-zero}).
The derivations given in the paper can be directly generalised on the case of
space with dimensionality larger than four.
According to eq.~\eqref{eq:a-6}, this requires the knowledge of corresponding
re-coupling coefficients.
Since the four-dimensional HSH are proportional to the wave functions of the
Hydrogen atom, various integrals involving those functions can be
calculated using the technique of multipole expansions developed above.
Examples of such calculations will be reported in a forthcoming publication.

Finally, we note that eq.~\eqref{eq:mult-exp-1} can be used for the derivation
of the general multipole expansion of the function
$f(r) C_j (\hr)$ where $\vvr=\vvr_1+\vvr_2$.
Namely, we write the Taylor series for the function $f(r)$,
\begin{equation}
  \label{eq:f-1}
  f(r) C_j (\hr) = \sum_{n=0}^\infty f_n r^n C_j (\hr).
\end{equation}
Each term of this equation can be decomposed using eq.~\eqref{eq:mult-exp-1},
which leads to the equation
\begin{equation}
  \label{eq:f-2}
  f(r) C_j (\hr) = \sum_{l,l'}
C^{(j)}_{ll'} (r_1,r_2) \;
\{ C_{l}(\hr_1) \otimes C_{l'} (\hr_2) \}_j,
\end{equation}
where the coefficients $C^{(j)}_{ll'}$ are defined by
\begin{equation}
  \label{eq:f-3}
C^{(j)}_{ll'} (r_1,r_2) = \sum_{n=0}^\infty f_n B^{(nj)}_{ll'}(r_1,r_2).
\end{equation}
The expressions for the parameters $B^{(nj)}_{ll'} (r_1,r_2)$ are given
by eq.~\eqref{eq:mult-exp-1}.

\acknowledgments

This work has been supported in part by the BRHE joint program of CRDF and
Russian ministry of education under the grant No.~Y2-CP-10-02.
I am grateful to Jan-Michael Rost for careful reading of the manuscript and
providing useful suggestions.


\appendix

\section{Expressions for four-dimensional Clebsch-Gordan coefficients}
\label{sec:expr-four-dimens}

The expression for $C$-type CGC simplifies for $j=j_1+j_2$.
In this case the $9j$-symbol in the definition (\ref{eq:clebsh-c}) of $C$-type
CGC can be re-written in terms of $3d$ CGC, see eq.~(9) of Sec.~10.8.3 of
\cite{Varsh}.
As a result, one arrives at the equation
\begin{multline}
  \label{eq:cg-sum}
C^{(j_1+j_2) \lambda \alpha}_{j_1 \lambda_1 \alpha_1;\,
 j_2 \lambda_2 \alpha_2}
=  C^{\lambda 0}_{\lambda_1 0\, \lambda_2 0}
C^{\lambda \alpha}_{\lambda_1 \alpha_1 \, \lambda_2 \alpha_2}
\frac{j_1 ! j_2 !}{(j_1+j_2)!}  \\
\times 
\left[
\frac{(j_1+j_2+\lambda+1)!\, (j_1+j_2-\lambda)!\, 
(2\lambda_1+1)\,(2\lambda_2+1)}{(j_1+\lambda_1+1)!\,
 (j_1 -\lambda_1)!\,(j_2+\lambda_2+1)!\, (j_2-\lambda_2)!\,(2\lambda+1)}
\right]^{1/2}.
\end{multline}
This equation simplifies significantly for $\lambda_1=\alpha_1=0$.
In this case the three-dimensional CGC are equal to unity, so that
\begin{equation}
  \label{eq:cg-sum-1}
C^{(j_1+j_2) \lambda \alpha}_{j_1 00;\,
 j_2 \lambda_2 \alpha_2}
= \delta_{\lambda,\lambda_2} \delta_{\alpha,\alpha_2}
\frac{j_2 !}{(j_1+j_2)!} 
\left[
\frac{(j_1+j_2+\lambda+1)!\, (j_1+j_2-\lambda)!}{(j_1+1)\,(j_2+\lambda+1)!\,
 (j_2-\lambda)!}
\right]^{1/2}.
\end{equation}
For $\lambda=\alpha=0$ this equation reduces to
\begin{equation}
  \label{eq:cg-sum-2}
 C^{(j_1+j_2) 00}_{j_1 00;\, j_2 00}
= \left[
\frac{j_1+j_2+1}{(j_1+1)\,(j_2+1)}
\right]^{1/2}
\end{equation}
which is in agreement with the general equation \eqref{eq:cgc-zero-proj}.
We consider also the particular case of eq.~(\ref{eq:cg-sum}) for
maximal values of projections $\lambda$,
\begin{equation}
  \label{eq:cg-sum-3}
  C^{(j_1+j_2)\, (j_1+j_2) \alpha}_{j_1 j_1 \alpha_1;\, j_2 j_2 \alpha_2}
= C^{(j_1+j_2) \alpha}_{j_1 \alpha_1\, j_2 \alpha_2}.
\end{equation}
For maximal possible values of both $\lambda$- and
$\alpha$-projections one has
\begin{equation}
  \label{eq:cg-4}
  C^{(j_1+j_2)\, (j_1+j_2)\, (j_1+j_2)}_{j_1 j_1 j_1;\, j_2 j_2 j_2} =1.
\end{equation}
We present also the expression for CGC with $j=j_2-j_1$, which can be derived
from (\ref{eq:cg-sum}) and the symmetry relations (\ref{eq:cg-1}) and
(\ref{eq:cg-2}),
\begin{multline}
  \label{eq:cg-dif-1}
C^{(j_2-j_1) \lambda \alpha}_{j_1 \lambda_1 \alpha_1;\,
 j_2 \lambda_2 \alpha_2}
=  C^{\lambda_2 0}_{\lambda 0\, \lambda_1 0}
C^{\lambda \alpha}_{\lambda_1 \alpha_1 \, \lambda_2 \alpha_2}
\frac{j_1 ! (j_2-j_1+1)!}{(j_2+1)!} \\
\times 
\left[
\frac{(j_2+\lambda_2+1)!\, (j_2-\lambda_2)!\, 
(2\lambda_1+1)}{(j_1+\lambda_1+1)!\,(j_1 -\lambda_1)!\,
(j_2-j_1+\lambda+1)!\, (j_2-j_1-\lambda)!} \right]^{1/2}.
\end{multline}

It is also of interest to consider CGC with some set of projections equal to
zero, e.g. $\lambda_1=\alpha_1=0$.
In this case, $\lambda=\lambda_2$ and $\alpha=\alpha_2$ and the
$9j$-coefficient in the definition \eqref{eq:clebsh-c} of CGC reduces to
the $6j$-coefficient \cite{Varsh},
\begin{equation}
  \label{eq:cgc-z-1}
  C^{j \lambda \alpha}_{j_1 0 0; j_2 \lambda \alpha} =
(-1)^{\lambda+(j_1+j_2+j)/2}\, \frac{j+1}{\sqrt{j_1+1}} \,
\left\{
  \begin{array}{ccc}
    \lambda & \frac{j}{2} & \frac{j}{2} \\
    \frac{j_1}{2} & \frac{j_2}{2} & \frac{j_2}{2}
  \end{array}
\right\}.
\end{equation}
In particular, for $j_1=1$ we have that $j=j_2\pm1$ and evaluating the
$6j$-symbol to its explicit form we obtain
\begin{equation}
  \label{eq:cgc-z-2}
      C^{(j-1) \lambda \alpha}_{1 0 0; j \lambda \alpha} = 
\frac{\sqrt{(j - \lambda) \,(j+\lambda+1)}}{ (j+1)\sqrt2}, \quad 
  C^{(j+1) \lambda \alpha}_{1 0 0; j \lambda \alpha} = 
\frac{\sqrt{(j - \lambda+1) \,(j+\lambda+2)}}{(j+1) \,\sqrt2}.
\end{equation}


\end{document}